\documentclass[12pt]{iopart}

\usepackage[cp1251]{inputenc}
\usepackage{iopams}
\usepackage{graphicx,epsfig}
\begin{document}

\title[]{Two gaps in the axiomatics of quantum mechanics}

\author{N. L. Chuprikov}

\address{Tomsk State Pedagogical University, 634041, Tomsk, Russia}
\ead{chnl@tspu.edu.ru} \vspace{10pt}

\begin{abstract}
This paper identifies and fixes two gaps in the axiomatics of modern quantum mechanics. The first arises from the
unjustified application of the superposition principle in modern quantum scattering theory. The second arises from the
incompleteness of the Born interpretation of the wave function.
\end{abstract}


\newcommand{\ppp}{\mbox{\hspace{5mm}}}
\newcommand{\ppa}{\mbox{\hspace{15mm}}}
\newcommand{\ppb}{\mbox{\hspace{20mm}}}
\newcommand{\ooo}{\mbox{\hspace{3mm}}}
\newcommand{\ooa}{\mbox{\hspace{1mm}}}

\section{Introduction} \label{int}

It is widely believed that the lack of a consistent physical interpretation of nonrelativistic quantum mechanics (QM) is
due to the fundamental incompatibility between the properties of the microworld and the laws of classical mechanics.
However, as will be shown below, this situation is explained by the fact that standard QM, as the Schr\"{o}dinger
representation (SR) of canonical commutation relations (CCR), contains only one postulate that adequately reflects the
properties of the microworld -- the Schr\"{o}dinger equation. It was this equation, introduced by Schr\"{o}dinger as a
postulate, that made QM such a successful theory for constructing mathematical models of quantum phenomena. As for its
remaining postulates and principles, for two reasons they distort the physical properties of the microworld, making a
consistent physical interpretation of QM impossible. In this regard, the analysis of the completeness and consistency of
the system of SR is a task of primary importance.

\section{Mathematical definition and axiomatics of SR} \label{int}

\subsection{Mathematical definition of SR and quantum phenomena it encompasses}

The mathematical definition of SR (see pp. 83 and 84 in \cite{Ara}) implies that the state vectors to satisfy the
Schr\"{o}dinger equation constitute a separable complex Hilbert vector space $\mathcal{H}=L^2(\mathbb{R}^n)$; for a
one-particle system $n=1,2,3$. The coordinate and momentum of a particle are represented in SR by the operators
$\hat{\mathbf{r}}$ and $\hat{\mathbf{p}}$, which are the closures of the multiplication operator $\mathbf{r}$ and the
differentiation operator $-i\hbar\mathbf{\nabla}$ in the space $C^\infty_0(\mathbb{R}^n)$ or in the Schwartz space
$\mathcal{S}(\mathbb{R}^n)$. Their components satisfy the Heisenberg CCR:
\begin{eqnarray*}
\fl [\hat{x}_i,\hat{x}_j]=0,\ooo [\hat{p}_i,\hat{p}_j]=0,\ooo [\hat{x}_i,\hat{p}_j]=i\hbar\delta_{ij}\ppp (i,j=1,\ldots
,n).
\end{eqnarray*}
Both these operators are unbounded.

SR based on the space $C^\infty_0(\mathbb{R}^n)$ covers single-particle quantum phenomena in which the particle moves in a
bounded spatial region (for example, in an infinitely deep potential well) and its energy spectrum is discrete. SR based
on the Schwarz space $\mathcal{S}(\mathbb{R}^n)$ covers phenomena in which the particle moves in the entire (simply
connected) configuration space $\mathbb{R}^n$ and therefore its motion is not compactly supported. This class includes
phenomena in which the probability of finding a particle at infinity is zero and its energy spectrum is discrete (an
example is the problem of a quantum harmonic oscillator); it also includes scattering phenomena in which the probability
of finding a particle at infinity is non-zero as $t\to\mp\infty$ and its energy spectrum is continuous.

\subsection{Axioms and physical principles of SR} \label{true}

At present, all three types of single-particle quantum phenomena are described by a single set of quantum-mechanical
axioms adapted to SR in \cite{Prug} (see pp. 269 and 292):

{\bf Axiom 1}. Any closed quantum-mechanical system $S$, which represents a nonrelativistic spinless quantum particle
moving in the configuration space $\mathbb{R}^n$ under the action of a static external field $V(\mathbf{r})$, is
associated with a time-independent separable complex Hilbert vector space $\mathcal{H}=L^2(\mathbb{R}^n)$, in which the
quantum mechanical theory of the system is formulated. At any time $t$ there is a ``ray'' in $\mathcal{H}$ -- the set of
unit vectors $|\psi\rangle_t$ defined up to a phase factor -- which describes a pure state of $S$ at the moment $t$. In
the coordinate representation, the state vector $|\psi\rangle_t$ represents the wave function $\psi(\mathbf{r},t)$.

{\it Comment}: The independence of the Hilbert space $\mathcal{H}$ on $t$ was introduced by von Neumann. On p. 102 of
\cite{Neu}, he writes: ``...since the Hilbert space is formed only with respect to the variables of the configuration
space (and normalization is applied only to them), the dependence on $t$ should not be taken into account here.''

{\bf Axiom 2}. In this theory, the position and momentum of a particle are represented by the operators $\hat{\mathbf{r}}$
and $\hat{\mathbf{p}}$, which are the closures of the multiplication operator $\mathbf{r}$ and the differentiation
operator $-i\hbar\mathbf{\nabla}$ defined in the space $C^\infty_0(\mathbb{R}^n)$ or in the Schwartz space
$\mathcal{S}(\mathbb{R}^n)$; their components satisfy the Heisenberg CCR:
\begin{eqnarray*}
\fl [\hat{x}_i,\hat{x}_j]=0,\ooo [\hat{p}_i,\hat{p}_j]=0,\ooo [\hat{x}_i,\hat{p}_j]=i\hbar\delta_{ij}\ppp (i,j=1,\ldots
,n).
\end{eqnarray*}
Any other observable $A$ is represented in this theory by a linear self-adjoint operator that is either a polynomial in
one of the operators $\hat{\mathbf{r}}$ and $\hat{\mathbf{p}}$, or a polynomial in the set of joint (mutually commuting)
components of these vector operators. The fact that $[\hat{x}_j,\hat{p}_j]\neq 0$ for $j=1,\ldots ,n$ means that different
experimental setups are required to measure these quantities for the particle in the pure state $|\psi\rangle_t$ (see,
i.e., p.82 in \cite{Ara}).

{\bf Axiom 3}. Each quantum-mechanical system $S$ is represented in the theory by a Hamiltonian
$\hat{H}=\hat{\mathbf{p}}^2/2m+V(\mathbf{r})$ -- a densely defined linear self-adjoint operator in $\mathcal{H}$.

{\it Comment}: $\hat{H}$ plays a dual role in the theory: it represents the total energy of the system $S$, as well as it
plays a key role in the equation of motion of this system.

{\bf Axiom 4}. The quantum dynamics of a closed system $S$ is described by the Schr\"{o}dinger equation
\begin{eqnarray}\label{2}
\fl i\hbar \frac{\partial |\psi\rangle_t}{\partial t}=\hat{H}|\psi\rangle_t \ppp \Rightarrow\ppp
i\hbar\frac{\partial\psi(\mathbf{r},t)}{\partial t}=\hat{H}\psi(\mathbf{r},t).
\end{eqnarray}

{\it Comment}: The Schr\"{o}dinger dynamics is unitary -- it preserves the norm of $|\psi\rangle_t$.

{\bf Axiom 5}. {\it Born's rule}: For any state $|\psi\rangle_t$, the quantity $|\psi(\mathbf{r},t)|^2$ gives the
probability density of detecting a particle at time $t$ in the infinitesimal neighborhood of the point $\mathbf{r}$. For
the particle in the state $|\psi\rangle_t$, the expectation value of the observable $A$ is determined by the scalar
product
\begin{equation} \label{1}
\fl \langle\psi|\hat{A}|\psi\rangle_t=\int_{\mathbb{R}^3}  \psi^*(\mathbf{r},t) \hat{A}\psi(\mathbf{r},t)d\mathbf{r}.
\end{equation}

{\bf Axiom 6} or {\it von Neumann's reduction postulate (wave function collapse)}: If $\mathbf{r}_0$ is the result of
measuring the coordinate of a particle in the state $\psi(\mathbf{r},t)$, then this means that the measurement process
instantly reduced this state to the eigenstate $\delta(\mathbf{r}-\mathbf{r}_0)$ of the coordinate operator, corresponding
to the eigenvalue $\mathbf{r}_0$.

It should be noted that von Neumann's postulate is not considered in \cite{Prug} as an axiom of SR, since it contradicts
axioms (1) and (4). At the same time, as will be shown later, the need to introduce the collapse postulate arose because
the system of axioms (1)–(5) for closed systems is internally inconsistent and, as a consequence, leads to the
``measurement problem''. Thus, although this von Neumann postulate is indeed alien to standard quantum mechanics, it can
be excluded only after identifying and correcting the errors in the system of axioms (1)–(5).

\section{On the false and true role of the superposition principle in SR} \label{superpose}

\subsection{The superposition principle as a postulate and as a theorem} \label{Weyl}

As is seen, the superposition principle is not contained in the system of quantum-mechanical axioms (1)-(5) presented in
the monograph for mathematicians \cite{Prug}. At the same time, this principle plays a key role in Dirac's axiomatics. For
physicists, the linearity of the Schr\"{o}dinger equation is a necessary and {\it sufficient} condition for the validity
of the superposition principle in all single-particle quantum phenomena described by the Schr\"{o}dinger equation. For
them, this principle is embedded in Axioms (1) and (4), and they consider the statement

{\it ``If $\psi_1(\mathbf{r},t)$ and $\psi_2(\mathbf{r},t)$ are two independent normalized solutions to the
Schr\"{o}dinger equation, then their linear combination $c_1\psi_1(\mathbf{r},t)+c_2\psi_2(\mathbf{r},t)$, where
$|c_1|^2+|c_2|^2=1$, is also a normalized solution to the Schr\"{o}dinger equation''}

\noindent as an equivalent to the statement

{\it ``If $\psi_1(\mathbf{r},t)$ and $\psi_2(\mathbf{r},t)$ are two pure states in which a particle can reside, then their
linear combination is $c_1\psi_1(\mathbf{r},t)+c_2\psi_2(\mathbf{r},t)$ represents another pure state it may be in''}.

\noindent The first statement represents the formulation of the superposition principle in the space of solutions of the
Schr\"{o}dinger equation, while the second extends this formulation onto the Hilbertian state space.

At the same time, mathematicians are more familiar with the theorem on irreducibility of SR (see pages 88 and 89 in
\cite{Ara}), which serves for them as a mathematical justification for the superposition principle. For them, this
principle is embedded in the very concept of the Hilbert space associated with the process under study, and the main task
consists only in proving the absence of superselection rules that can limit the definition domain of observables'
operators in this space. As is stressed in \cite{Prug} on p. 301, ``{\it The natural question to ask is whether all the
vectors in the Hilbert space $\mathcal{H}$ assigned to the system [Axiom 1] are state vectors.}''

One of the most recent proofs of this theorem was presented on pages 87 and 88 of \cite{Ara}, and like others, it is based
on the Weyl form of the CCR. The main intrigue underlying all existing proofs of this theorem is that SR is related to
Heisenberg's CCR, but not to Weyl's CCR. And all monographs on quantum mechanics (see, for example, \cite{Ara,Prug,Re1})
emphasize the fact that these two representations are only formally equivalent, since the position and momentum operators
satisfying Heisenberg's CCR are unbounded.

Nevertheless, it is generally accepted that in the case of quantum phenomena with a single spinless nonrelativistic
particle, the validity of the superposition principle is unquestionable, and superselection rules should not arise here.
As pointed out on page 302 in \cite{Prug}, the reason is that ``... the [presently known] superselection rules are
primarily of significance only in theories treating systems with a variable number of particles, for instance in field
theories... Since we are primarily concerned with nonrelativistic quantum mechanics, in which the nature of the particles
and their number does not change, the existence of the above-mentioned superselection rules will not be of any immediate
concern to us''.

That is, the unboundedness of the position and momentum operators in SR is currently considered as a mathematical property
that does not have any decisive physical significance. As emphasized in \cite{Str} on p. 58, ``{\it The physical reason
for this mathematical obstruction is that, strictly speaking, $q$ and $p$ are not observables in the operational sense...;
due to the scale bounds of experimental apparatuses, one actually measures only bounded functions of q and p, (namely the
position inside the volume accessible by the experimental apparatus and the momentum inside an interval given by the
energy bounds set by the apparatus). Thus, a formulation based on the Heisenberg algebra involves an (in fact physically
harmless) extrapolation with respect to the operational definition of observables}''. As a result, the irreducibility of
SR is now considered an indisputable fact.

However, the properties of the state space of a particle participating in the quantum process under study, as well as the
properties of operators acting in this space are determined by the Schr\"{o}dinger equation with the corresponding
potential $V(\mathbf{r})$. If the external field is such that the energy spectrum of a particle is discrete, and therefore
the probability of detecting it at infinity is zero, then the unboundedness of the position operator can be neglected. As
regards scattering phenomena, when the particle is by definition at infinity for $t\to -\infty$ and/or $t\to +\infty$ (see
p. 12 in \cite{Re3}), the unboundedness of the position operator cannot be ignored. Therefore, the SR irreducibility
theorem is inapplicable to scattering phenomena -- the current practice of treating scattering states as pure states lacks
a rigorous mathematical basis.

\subsection{On the Hilbert state-space associated with scattering a particle on a one-dimensional $\delta$-potential} \label{delta}

Volovich's words, spoken in connection with quantum mechanics itself (see \cite{Vol}), are quite applicable to the current
situation in modern quantum scattering theory: {\it ``There is a gap between an abstract approach to the foundations and
the very successful pragmatic approach to quantum mechanics which is essentially reduced to the solution of the
Schr\"{o}dinger equation. If we will be able to fill this gap then perhaps it will be possible to get a progress in the
investigations of foundations because in fact the study of solutions of the Schr\"{o}dinger equation led to the deepest
and greatest achievements of quantum mechanics...}''

Since the application of the superposition principle to scattering problems has no solid basis, any scattering process in
quantum mechanics can only be reliably studied based on solutions of the corresponding Schr\"{o}dinger equation. This is
clearly demonstrated in the article \cite{Ch1}, which presents an exact quantum-mechanical model of scattering a particle
on a one-dimensional $\delta$-potential. This is the first scattering model based on an analysis of the asymptotes of
nonstationary solutions of the corresponding Schr\"{o}dinger equation in the configuration space $\mathbb{R}$ as
$t\to\mp\infty$.

From this model it follows that the standard formulation of the superposition principle and the theorem on the
irreducibility of SR contradict the asymptotic properties of non-stationary solutions of the Schr\"{o}dinger equation in
this scattering problem. Axioms (1)-(3) are invalid, since the asymptotes of the scattering states form, as
$t\to\mp\infty$, a reducible Hilbert space, consisting of two superselection sectors invariant under the action of the
observables' operators. That is, these operators, including the Hamilton operator, are defined not on the entire
$\mathcal{H}$, but only in its superselection sectors. When $t\to \infty$, any scattering state, as being a superposition
of transmitted and reflected wave packets belonging to two different superselection sectors, is a mixed quantum state.
Calculating the expectation values of any observable for these scattering state has no physical sense, because he unitary
Schr\"{o}dinger dynamics in this scattering process intersects the boundaries of superselection sectors. Thus, the
linearity of the Schr\"{o}dinger equation is only a necessary, but not sufficient, condition for the validity of the
superposition principle in scattering problems.

\subsection{The Schr\"{o}dinger cat paradox as an indication for a revision of the quantum mechanical model of radioactive
nucleus decay} \label{cat}

We are now ready to take a fresh look at Schr\"{o}dinger's cat thought experiment in which the superposition of
microscopically distinct states ``undecayed nucleus'' and ``decayed nucleus'' transforms into the superposition of
macroscopically distinct states ``alive cat'' and ``died cat''. The paradox associated with this experiment is that the
second superposition, like the first, should be considered a pure state in quantum mechanics (see axioms). At the same
time, no one has ever observed such superpositions—a cat, as a macroscopic object, is always either alive or dead.

Since no one doubted the applicability of the superposition principle to the decay of a radioactive nucleus, this
experiment came to be viewed as a measurement problem or a problem of macroobjectification (the latter term is preferable
in the context of our study). To address this problem, the wave function collapse postulate was introduced, according to
which the superposition of microscopically distinct states of the radioactive nucleus decoheres (is transformed) into a
mixture of macroscopically distinct states of the cat due to the irremovable influence of the environment -- the state of
a quantum system becomes entangled with the environment's state.

However, this ``solution'' to the cat paradox distracts us from the fact that the quantum-mechanical model of radioactive
nucleus decay used by Schr\"{o}dinger in his thought experiment is purely speculative. It is not based on the
time-dependent solutions of the corresponding Schr\"{o}dinger equation and the properties of their asymptotes as
$t\to\mp\infty$. It is based on the {\it assumption} that the linearity of the Schr\"{o}dinger equation is a {\it
sufficient} condition for the validity of the superposition principle in all scattering problems described by this
equation.

The modern quantum-mechanical model of particle scattering on a one-dimensional $\delta$-potential is also based on this
assumption. And, as shown in \cite{Ch1}, this model is incorrect, since the asymptotes of the scattering states form a
reducible Hilbert space as $t\to \mp\infty$ -- the action of the superposition principle in this space is limited to
superselection sectors. Thus, to adequately interpret Schr\"{o}dinger's thought experiment with a cat, it is necessary to
develop an accurate model of the decay of a radioactive nucleus which would be based on the properties of the
nonstationary solutions of the corresponding Schr\"{o}dinger equation at $t\to \infty$.

However, in this thought experiment, the radioactive atom can be replaced by a particle scattering on a
(quasi)one-dimensional delta potential -- the cat is in a mixed state, just like the particle in the final stage of
scattering. Therefore, even now, that is, before a precise model of the radioactive nucleus is developed, it can be argued
that the Schr\"{o}dinger cat paradox is not a macroobjectification problem.

This paradox points to a serious flaw in the foundations of modern quantum mechanics, which lies in the unjustified
extension of the superposition principle and the SR irreducibility theorem to scattering processes involving the decay of
a radioactive nucleus. This means that there is currently no rigorously grounded quantum theory of scattering. To fill
this gap, it is necessary to reexamine quantum mechanical models of scattering processes to determine whether they contain
asymptotic superselection rules. This means that each scattering model must be revised taking into account the properties
of the non-stationary solutions of the corresponding Schr\"{o}dinger equation as $t\to \mp\infty$.

\section{On the true (complete) interpretation of the wave function in SR} \label{Born}

\hspace*{\parindent} So, we have established that the current formulation of the superposition principle as a universal
principle of quantum mechanics applicable to all scattering phenomena is erroneous. This means, there is a gap in the
foundations of modern quantum mechanics—the modern nonstationary quantum scattering theory, which relies unconditionally
on the superposition principle, must be reconsidered.

Another gap in its foundations arises in the description of one-particle phenomena with a discrete energy spectrum, when
the validity of the superposition principle and the irreducibility theorem of SR are beyond doubt. The essence of this
problem is that Born's statistical interpretation of the wave function describing a pure state is incomplete -- it reveals
the physical meaning of the squared modulus of the wave function (Axiom 5), but completely ignores its phase. According to
this interpretation, the wave function that defines the pure state of a particle does not impose any restrictions on the
values of observables –- it imposes only statistical restrictions on the observables' values.

Moreover, the statistical interpretation of the squared modulus of the wave function can be understood in two ways. For
example, the (majority) of eminent physicists (including Bohr) interpret the function $|\psi(\mathbf{r},t)|^2$ as the
probability density of detecting a particle in an infinitesimal neighborhood of the point $\mathbf{r}$ at time $t$.
Another group (including Ballentine) interpret it as the distribution function of (non-interacting) particles in the
quantum statistical ensemble over points in the space $\mathbb{R}^3$ at time $t$.

According to Bohr, the wave function provides a complete description of the state of a particle. It predicts the
probabilities of the occurrence of values of observables during their measurement, as well as their average values.
According to Bohr, these values do not exist before measurement; they appear during measurement. In particular, the
process of ``acquiring'' a certain position by a particle in space $\mathbb{R}^3$ during the measurement of the particle's
coordinate is governed by axiom 6. That is, as in the interpretation of the thought experiment with the cat, in Bohr's
interpretation of the wave function (as a pure state representing a superposition of eigenstates of the particle's
coordinate operator) the measurement problem arises. To distinguish it from the above-mentioned problem of
macroobjectification, we call it the problem of ``eigenvalue actualization'' (see also p. 77 in \cite{Jae}).

Although the origins of the macroobjectification problem and the "eigenvalue actualization" problem are different, the
very fact that both appear in quantum mechanics is unacceptable, since Axiom 6 is incompatible with axioms (1)-(4). And,
more importantly, neither is a problem of quantum mechanics. The first arose from an erroneous interpretation of the cat
thought experiment, and the second is the result of an erroneous interpretation of the wave function describing a pure
state—this wave function defines the state not of a single particle, but of the corresponding one-particle quantum
ensemble.

A statistical interpretation of the wave function based on the concept of a quantum statistical ensemble is presented in
Ballentine's paper \cite{Bal} (see also references to Blokhintsev and other proponents of this interpretation). According
to Ballentine, the transition to an ensemble interpretation of the wave function automatically solves the measurement
problem (it is important to emphasize that the measurement problem is not currently divided, as in our approach, into the
problem of macroobjectification and the problem of eigenvalue actualization; this is due to the fact that any
single-particle state that is defined by a wave function satisfying the Schr\"{o}dinger equation is considered a pure
state). However, this is not the case—the transition to an ensemble interpretation automatically solves only the
eigenvalue actualization problem that arises for pure states.

Another drawback of the ensemble interpretation \cite{Bal} is that the Born interpretation does not contain any
indications regarding other physical properties of the ensemble particles, other than that the ensemble particles are
distributed over points in space $\mathbb{R}^3$ with probability $|\psi(\mathbf{r},t)|^2$. In this regard, attempts have
been made to develop a deeper interpretation of the wave function that would reveal the physical meaning of the phase of
the wave function. First and foremost, it is worth mentioning Bohm's quantum mechanics \cite{Bohm}.

In this approach, the phase $S(\mathbf{r},t)/\hbar$ of the wave function $\psi(\mathbf{r},t)=\sqrt{w(\mathbf{r},t)}\ooa
e^{i\frac{S(\mathbf{r},t)}{\hbar}}$ determines in the space $\mathbb{R}^3$ the particle momentum field
$\mathbf{p}(\mathbf{r},t)=\mathbf{\nabla}S(\mathbf{r},t)$. Together with the distribution function
$w(\mathbf{r},t)=|\psi(\mathbf{r},t)|^2$, this field satisfies the continuity equation that describes an incompressible
``fluid''. As a consequence, a single-particle quantum ensemble is considered an incompressible ``fluid'' whose
streamlines represent particle trajectories. Particle motion along these trajectories occurs under the influence of the
external field $V(\mathbf{r})$ and the ``quantum potential'' $U_R(\mathbf{r},t)$:
\begin{eqnarray}\label{4}
\fl U_R= K_w+U_w;\ppp K_w=\frac{\hbar^2}{8m}\left(\frac{\mathbf{\nabla}w}{w}\right)^2,\ooo U_w= -
\frac{\hbar^2}{4m}\frac{\mathbf{\nabla}^2 w}{w}.
\end{eqnarray}
Thus, if the initial position of an individual particle in the ensemble is known, its position and momentum can be
predicted for any subsequent instant. This means that in Bohm's quantum mechanics, both observable quantities are
classified as hidden variables. As consequence, the validity of Bohm's approach and his version of the quantum statistical
ensemble has been questioned by much of the scientific community.

Another version proposed was the model of Brownian motion \cite{Nels}. Its advantage over Bohm's model is that the
particles involved in Brownian motion, although possess trajectories, their coordinate and momentum are unpredictable and
hence are not hidden variables. However, the mere fact that the equation describing randomly colliding Brownian particles
coincides with the one-particle Schr\"{o}dinger equation is not sufficient grounds for considering a quantum ensemble of
non-interacting particles as a set of Brownian particles. Therefore, Nelson's approach is also considered incompatible
with the axioms of quantum mechanics.

Thus, there currently exists no consistent concept of a quantum statistical ensemble with rich physical content and free
of hidden variables. In this regard, we turn to the recent work \cite{Ch2} and \cite{Chu}, where the properties of a pure
ensemble, whose state is described by a wave function satisfying the Schr\"{o}dinger equation, are examined in details.
This approach, unlike Bohmian mechanics, provides an interpretation of the wave function phase without introducing an
additional postulate. Apart from the Schr\"{o}dinger equation, written as two real equations for the modulus and phase of
the wave function, a significant role is played by the formula (\ref{1}) from Axiom 5 for determining the expected values
of observables.

The basic idea of this approach is that if for any operator $\hat{A}$ that does not commute with the position operator,
the integral in (\ref{1}) has a physical meaning, then its integrand must also have a physical meaning. In the coordinate
representation, the mathematical expectation of the observable $A$ in the state $\psi(\mathbf{r},t)$ can be written in
terms of the field $\hat{A}(\mathbf{r},t)$ of this operator:
\begin{equation} \label{5}
\fl \langle\psi|\hat{A}|\psi\rangle=\int_{\mathbb{R}^3} \psi^*(\mathbf{r},t) \hat{A}\psi(\mathbf{r},t)d\mathbf{r}\equiv
\int_{\mathbb{R}^3} A(\mathbf{r},t)w(\mathbf{r},t)d\mathbf{r},
\end{equation}
where $A(\mathbf{r},t)=\mathrm{\Re}[\psi^*(\mathbf{r},t)\hat{A}\psi(\mathbf{r},t)]/w(\mathbf{r},t)$.

For the operators of the particle's momentum, its kinetic and total energy, which are included in the Schr\"{o}dinger
equation, these fields are determined uniquely:
\begin{eqnarray*}
\fl \mathbf{p}(\mathbf{r},t)=\mathrm{\Re}\left[\psi^*(\mathbf{r},t)\hat{\mathbf{p}}\psi(\mathbf{r},t)\right]/w(\mathbf{r},t)
= \mathbf{\nabla}S(\mathbf{r},t),\\
\fl E_{kin}(\mathbf{r},t)=
\Re\left[\psi^*(\mathbf{r},t)\frac{\hat{\mathbf{p}}^2}{2m}\psi(\mathbf{r},t)\right]\Big/w(\mathbf{r},t)
=\frac{\mathbf{p}^2(\mathbf{r},t)}{2m}+U_R(\mathbf{r},t)=\frac{\mathbf{p}^2}{2m}+K_w+U_w;\\
\fl E(\mathbf{r},t)=
\mathrm{\Re}[\psi^*(\mathbf{r},t)\hat{H}\psi(\mathbf{r},t)]/w(\mathbf{r},t)=E_{kin}(\mathbf{r},t)+V(\mathbf{r}).
\end{eqnarray*}
As is seen, the function $U_R(\mathbf{r},t)$, which Bohm regarded as a "quantum potential," is actually part of the field
of the particle's kinetic energy operator. The function $\mathbf{p}(\mathbf{r},t)$ cannot be interpreted as the particle's
momentum, since $\hat{\mathbf{p}}^2/2m$ is only one contribution to the field of the kinetic energy operator.

From the obtained expressions for the fields it follows that the wave function describing the state of a pure ensemble
determines at each point of the configuration space $\mathbb{R}^3$ at each moment of time not one value of the momentum
(as is assumed in Bohm mechanics), but two -- $\mathbf{p}_1(\mathbf{r},t)$ and $\mathbf{p}_2(\mathbf{r},t)$. And these
values are such that the expectation values of the particle's momentum and its kinetic energy can be represented in the
form
\begin{equation*}
\fl \int_{\mathbb{R}^3}  \mathbf{p}(\mathbf{r},t)w(\mathbf{r},t)d\mathbf{r}=\frac{1}{2}\int_{\mathbb{R}^3}
\left[\mathbf{p}_1(\mathbf{r},t)+\mathbf{p}_2(\mathbf{r},t)\right]w(\mathbf{r},t)d\mathbf{r},
\end{equation*}
\begin{equation*}
\fl\int_{\mathbb{R}^3} E_{kin}(\mathbf{r},t)w(\mathbf{r},t)d\mathbf{r}=\frac{1}{2}\int_{\mathbb{R}^3}
\left[\frac{\mathbf{p}_1^2(\mathbf{r},t)}{2m}+\frac{\mathbf{p}_2^2(\mathbf{r},t)}{2m}\right]w(\mathbf{r},t)d\mathbf{r}.
\end{equation*}
As shown in \cite{Ch2}, these conditions, as well as the restrictions on the operator fields that follow from the
Schr\"{o}dinger equation and its differential consequences, are satisfied by the following two fields of momentum values:
\begin{eqnarray}\label{6}
\fl \mathbf{p}_1(\mathbf{r},t)=\mathbf{p}(\mathbf{r},t)-\mathbf{p}_w(\mathbf{r},t),\ppp
\mathbf{p}_2(\mathbf{r},t)=\mathbf{p}(\mathbf{r},t)+\mathbf{p}_w(\mathbf{r},t);\ppp
\mathbf{p}_w=\frac{\hbar}{2}\frac{\mathbf{\nabla}w}{w}.
\end{eqnarray}
(Note that the function $U_w(\mathbf{r},t)$, although part of the field of the kinetic energy operator, does not enter
these expressions, since it makes zero contribution to the expected value of the kinetic energy of the particle.)

An important feature of the fields $\mathbf{p}_1(\mathbf{r},t)$ and $\mathbf{p}_2(\mathbf{r},t)$ is that, firstly, they
satisfy the Heisenberg inequalities, which impose restrictions on the variances of the particle's position and momentum;
secondly, the velocities $\mathbf{p}_1(\mathbf{r},t)/m$ and $\mathbf{p}_2(\mathbf{r},t)/m$ coincide with the velocities
$\mathbf{b}(\mathbf{r},t)$ and $\mathbf{b}_*(\mathbf{r},t)$ in Nelson's model \cite{Nels}, which he introduces to
characterize the motion of Brownian particles (without friction) forward and backward in time, respectively. That is, our
approach, developed strictly within the framework of the standard quantum-mechanical formalism, shows that these
velocities characterize not only Brownian motion, but also the dynamics of a single-particle quantum ensemble, thereby
confirming the dual nature of Schr\"{o}dinger dynamics established by Nelson.

At first glance, such duality is impossible, since the particles of a quantum ensemble do not interact with each other.
However, if we consider the fact that quantum particles are in principle indistinguishable from one another, then the
scenario of an imagin collision of non-interacting particles in a quantum ensemble is indistinguishable from the scenario
of a collision of Brownian particles. And since this scenario of a ``collision'' of non-interacting, indistinguishable
quantum particles occurs at every point in the configuration space $\mathbb{R}^3$, the dynamics of a single-particle
quantum ensemble as a whole turns out to be dual -- it is simultaneously deterministic and stochastic. It is important to
emphasize that the past and future of (identical) particles in an ensemble cannot be determined in principle. Therefore,
the position and momentum of a particle in this model of a pure ensemble are not hidden variables.

Finally, it is worth noting that this quantum ensemble model has some common features with the pre-quantum probabilistic
approach \cite{Khre}, based on the Kolmogorov concept of probability.

\section*{Conclusion}

Thus, we come to the conclusion that in quantum mechanics in the Schr\"{o}dinger representation one should distinguish the
following two fundamentally different classes of single-particle quantum phenomena: (a) phenomena in which the particle
energy spectrum is discrete; (b) phenomena in which its spectrum is continuous – scattering phenomena. In the case of the
first class of phenomena, axioms (1)–(4) remain unchanged, and the new version of Axiom 5 (Born's rule) should indicate
the existence in $\mathbb{R}^3$ of not only a probability field, but also fields of observables. In the case of scattering
phenomena, axioms (1)–(5) should be reformulated for non-separable Hilbert spaces in which asymptotic superselection rules
act. As for Axiom 6, it should be excluded in both cases.

\end{document}